# Fluids flow in granular aggregate packings reconstructed by high-energy X-ray computed tomography and lattice Boltzmann method

Qifeng Lyu [1,a], Anguo Chen [1], Jie Jia [1], Amardeep Singh [2], Pengfei Dai [1]

[1] Department of Civil Engineering, Changzhou University, Changzhou, Jiangsu 213164, China

[2] School of Civil Engineering and Architecture, Changzhou Institute of Technology, Changzhou, Jiangsu 213032, China

[a] Author to whom correspondence should be addressed: lqf@cczu.edu.cn

**Abstract**: Properties of fluids flow in granular aggregates are important for the design of pervious infrastructures used to alleviate urban water-logging problems. Here in this work, five groups of aggregates packing with similar average porosities but varying particle sizes were scanned by a high-energy X-ray computed tomography (X-CT) facility. The structures of the packings were reconstructed. Porosities were calculated and compared with those measured by the volume and mass of infilled water in the packing. Then pore networks were extracted and analyzed. Simulations of fluids flow in the packings were performed by using a lattice Boltzmann method (LBM) with BGK (Bhatnagar-Gross-Krook) collision model in the pore-network domain of the packings. Results showed wall effect on the porosity of aggregates packing was significant and the influence increased with the aggregate sizes. In addition, Poisson law and power law can be used to fit the coordination number and coordination volume of the packing's pore network, respectively. Moreover, the mass flow rates of fluids in the aggregates were affected by the porosities. On the two-dimensional slices, the mass flow rate decreased when the slice porosity increased. But for the three-dimensional blocks, the average mass flow rate increased with the volume porosity. And the permeability of the aggregates packing showed correlating change trend with the average pore diameter and fitting parameters of coordination volumes, when the sizes of aggregates changed. Though the limitation of merging interfaces causing fluctuation and discontinuity on micro parameters of fluid flow existed, the methods and results here may provide knowledge and insights for numerical simulations and optimal design of aggregate-based materials.

# 1 Introduction

Urban construction needs great quantities of infrastructures made of cementitious or asphaltic concrete [1], for instance, the pavement of road and highway, which is usually impermeable and thus may lead to water-logging on rainy days. To alleviate this problem, a concept of sponge city had been proposed which requires great quantities of pervious concrete and permeable pavement to drain off or store water [2, 3]. Usually, the coarse aggregates, i.e., gravels with sizes of 4.75~37.5 mm, are the main component of pervious pavement. Indeed, the base layer of the pervious pavement is directly made of granular aggregates and the top layer is cemented aggregates [1, 3]. Thus, understanding the fluids flow properties in the granular aggregates is crucial for an optimal design of the related infrastructures [2].

Fluids flow in the granular aggregates can be investigated by experiments on the tests of transport parameters like permeability [4, 5] which characterizes the hydrodynamic properties of the mass flow in the porous media [6]. The permeability can be measured by Darcy test which uses two constant pressure heads of a fluid to generate steady flow in the porous media and then the permeability can be calculated based on the flow rate, pressure gradient and sample sizes [5]. For convenience, the falling-head method can also be used in the granular packing though improvements are needed for the high flow rate of larger-size aggregates packing [7]. Moreover, analytical models correlating the macroscopic parameters like permeability, porosity and tortuosity of the porous media also exist [8, 9]. The microscopic parameters like aggregate shape and size may also influence the flow characteristics [10], which are usually investigated by numerical simulations [11]. Generally, two kinds of viewpoints were used in the simulations [12, 13], i.e., Eulerian viewpoint [14-16] like those used in finite element method and Lagrangian viewpoint [17-19] used in discrete element method. The former directly computes macroscopic properties of fluids whereas the latter is based on a microscopic view and thus is time consuming. Among the methods of fluid simulations, the lattice Boltzmann method [20, 21] which is a mesoscopic Lagrangian method has the merit of excellent efficiency in computing time and dealing with complex boundaries like those of gravel aggregates. On the other hand, to avoid the complex boundaries, the aggregates can be simulated by spheres [10, 22], and may also be simulated as irregular shapes by gluing multiple spheres [23], which requires more knowledge on the realistic shapes of aggregates [24]. To this end, the X-ray computed tomography (X-CT) can help a lot [25], though traditional X rays may not fully penetrate the macroscopic large-size packing. Thus, a high-energy X-CT is relatively more helpful.

For the granular aggregates used in infrastructures construction, understanding the fluids flow in their inner structures is crucial for optimal and intelligent designs of related porous materials. However, experiments spanning macro to micro scales are usually difficult, and the aggregate information required by numerical simulation is also hard to measure by traditional methods due to large quantities and densities of the aggregates. In this work, a high-energy industrial X-CT was used to scan five granular aggregates packings, and their pore structures and distributions were extracted and analyzed. In addition, larger packings were also created by merging the scanned and reconstructed digital structures. Based on these data, numerical simulation of fluids flow was conducted and analyzed. Though the limitation of merging interfaces causing fluctuation and discontinuity on micro parameters of fluid flow existed, the methods and results here may provide knowledge and insights for optimal design of aggregate-based porous materials.

## 2 Materials and Methods

### 2.1 Materials

Five groups of granular aggregates were packed and used in the experiments. The packing specimens are shown in Fig. 1 and the aggregate sizes of the packings are listed in Table 1. The aggregates used in this work are basalt, which were sieved by vibrating sieves with the grading mesh sizes listed in Table 1, i.e., 4.75 mm, 9.5 mm, 13.2 mm, 19 mm, 26.5 mm, 31.5 mm, 37.5 mm.

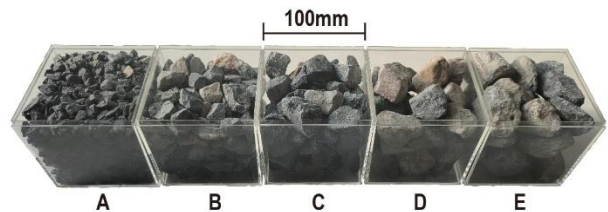

Figure 1. Granular aggregates packings.

Table 1. Particle sizes of the five granular aggregates packings.

| Packing | A | B | C | D | E |
| --- | --- | --- | --- | --- | --- |
| Size (mm) | 4.75~9.5 | 13.2~19 | 19~26.5 | 26.5~31.5 | 31.5~37.5 |

### 2.2 X-CT

The inner structure of the aggregates packing was detected by an industrial X-ray CT facility shown in Fig. 2. The CT facility consists of three parts which are X-ray source, sample table and X-ray detector. This is a high-energy X-CT which uses 450 kV voltage to generate strong X-rays with a maximum power of 4.5 kW that can easily penetrate 5 cm steel whose density is around 7~8 g/cm³. For the aggregate with a density around 2~3 g/cm³ (1/3 of that of steel), the X rays generated by this machine can fully penetrate the packing samples with 10 cm thickness.

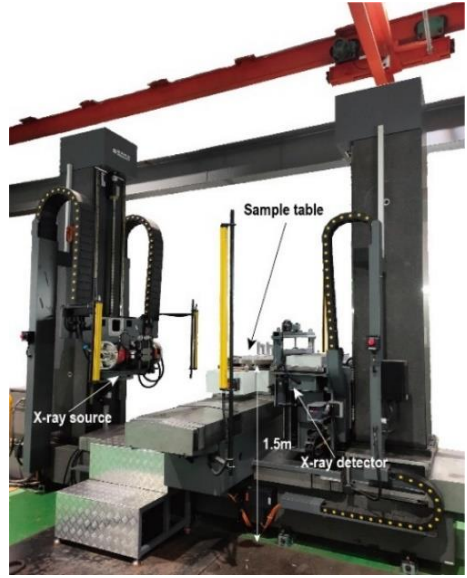

Figure 2. X-CT facility.

During a scan, X rays radiate from the source with an incident intensity $I_0$, then penetrate the aggregates packing on the sample table, and last is received by the detector with an attenuated X-ray intensity $I$ which follows the Beer-Lambert law:

$$I = I_0 e^{-\zeta x}, \tag{1}$$

where $e$ is the Euler's number, $\zeta$ is an attenuation factor affected by material densities, and $x$ is the packing's thickness in the direction of X-ray transmission. Thus, the intensity would reflect material densities varying in space. And at last, the varying intensities are stored as 32-bit floating-point data for 2048 × 2048 pixels on each slice with a resolution of 0.098 mm/pixel.

The CT images were processed by VG Studio and Avizo on a workstation with 128 GB DDR4 memory, NIVIDIA GeForce RTX3080 10GB GPU and Intel Xeon Silver-4210 × 2 CPU processors. Median filter was used to de-noise the raw images. This filter uses the neighbor's median to replace the value on each voxel. It has the merit of smoothing the images but without losing boundary information [26]. Then watershed segmentation method was used to classify the voxels into 2 groups, i.e., 0 and 1 which represent void and aggregate, respectively. The watershed segmentation is based on the simulation of water rises from a set of phase markers which represent different phases (e.g., void and aggregate) in the CT image. This method is more accurate than the traditional factorization segmentation method [26, 27]. After segmentation, the fraction of phases (void and aggregate) can be calculated based on the values of voxels. And indeed, the fraction of void is the porosity ($\phi$) whereas the fraction of aggregate is the density ($1 - \phi$) of the aggregates packing.

**2.3 Pore networks**

Based on the segmented images, pore networks can be extracted, which use spheres and tubes to represent pores and throats in the porous media. And the networks can be used to analyze the pore and throat diameters, mutual connections, coordination numbers etc. In this work, the code PNextract based on the maximal-ball algorithm [28] was used to extract the pore networks of the aggregates packing. This code was widely used [29], and the coordination number, pore and throat sizes, porosity calculated by this code agree well with experimental data [30]. OpenPNM [31] was used to transform the data structure of PNextract into the traditional network form used in NetworkX. Here OpenPNM and NetworkX are Python packages for network analyses.

Pore networks are similar to but different with social networks because the pore networks have volume whereas the social networks do not. Here the social networks mean connections of social individuals, like a person, group, or community, for instance, the networks of a person's friends, colleagues, or customers. Because of the difference, a coordination volume was proposed to characterize pore networks [26], which can be written as

$$c_v = V_{pi} + \sum_{j=1}^{n} V_{tj}, \tag{2}$$

where $V_{pi}$ represents the volume of pore $i$ in the pore network, and this pore has $n$ throats linking with it, $V_{tj}$ represents the volume of throat $j$ which is connected to pore $i$. And correspondingly, the coordination number was written as $c_n$.

In this work, the distribution of the coordination number and coordination volume were studied, and they were fitted by Poisson law and power law shown as:

$$f = \frac{e^{-\lambda} \lambda^{c_n}}{c_n!}, \tag{3}$$

$$f = \gamma c_v^{-\alpha}, \tag{4}$$

respectively, where $f$ represents the frequency of the variables, $\lambda, \gamma, \alpha$ are fitting parameters. Usually, the power law (Eq. 4) is used in logarithm coordinates system, where it can be transformed

into a straight line:

$$y = -\alpha x + \beta, \tag{5}$$

with $y = \log f$, $\beta = \log \gamma$, $x = \log c_v$.

Levenburg-Marquardt algorithm in SciPy (Python package) was used to fit the distributions of coordination number and coordination volume. And the coefficient of determination, $R^2$, was used to represent the goodness of fit, which can be written as

$$R^2 = 1 - \frac{\sum(y - \hat{y})^2}{\sum(y - \bar{y})^2}, \tag{6}$$

where $y$ represents the sample data, $\hat{y}$ represents the fitted data, and $\hat{y}$ represents the mean of the sample data. when $R^2$ approaches to 1, the fitting result is better.

**2.4 Numerical methods**

Stokes flow was used to simulate the water flow in the aggregates packing, which is simplified from the Navier-Stokes equation for the incompressible Newtonian fluid at low Reynolds numbers (i.e., laminar steady state) and can be described by the following equations [32]:

$$\nabla \cdot \mathbf{v} = 0, \tag{7}$$

$$\mu \nabla^2 \mathbf{v} - \nabla P = 0, \tag{8}$$

where $\nabla = \partial/\partial \mathbf{x}$ is the vector differential operator, in which $\mathbf{x}$ represents the coordinates vector in the 3-dimensional space, i.e., $x, y, z$; $\mathbf{v}$ is the velocity (vector) of the fluid; $\mu$ is the dynamic viscosity of the fluid and 0.001 Pa·s for water was used; $\nabla^2 = \partial^2/\partial x^2 + \partial^2/\partial y^2 + \partial^2/\partial z^2$ is the Laplacian operator; and $P$ is the pressure of the fluid.

The simulation was performed in mesoscopic scales by using a lattice Boltzmann method (LBM) with BGK (Bhatnagar-Gross-Krook) collision model [21, 33] in the pore-network domain of the aggregates. The equation of this method can be expressed as

$$f_i(\mathbf{x} + \mathbf{c_i}\Delta t, t + \Delta t) = f_i(\mathbf{x}, t) - \frac{1}{\tau}\left(f_i(\mathbf{x}, t) - f_i^{eq}(\mathbf{x}, t)\right) \tag{9}$$

where $f_i$ is the density distribution function for the flowing particles with velocity $\mathbf{c_i}$ at position $\mathbf{x}$ and time $t$ in the $i$ direction of the lattice, $\Delta t$ is the time step of the simulation, $\tau$ is the relaxation time, and $f_i^{eq}$ is the equilibrium distribution function which can be written as

$$f_i^{eq} = \rho \omega_i \left[1 + \frac{\mathbf{v} \cdot \mathbf{c_i}}{c_s^2} + \frac{(\mathbf{v} \cdot \mathbf{c_i})^2}{2c_s^4} - \frac{\mathbf{v} \cdot \mathbf{v}}{2c_s^2}\right] \tag{10}$$

where $\rho$ is the fluid density, $\omega_i$ is the weight of $i$ direction in the lattice, and $c_s$ is the speed of sound. The macroscopic properties (e.g., density and velocity) of the fluid can be obtained by

$$\rho = \sum_i f_i \tag{11}$$

$$\mathbf{v} = \frac{1}{\rho}\sum_i f_i \mathbf{c_i} \tag{12}$$

In the simulation, the no-slip boundary condition at fluid-solid interface was employed, and the sample's boundary was simulated by static pressure zones with a pressure difference of 0.1 Pa at the inlet and outlet surfaces of the packing whereas the other surfaces were sealed. The simulation mainly involves two steps, i.e., collision and streaming. After the simulation reaches to convergency, the permeability can be obtained by the Darcy's law [5, 34]:

$$k = \frac{Q\mu L}{A\,\Delta P}, \tag{13}$$

where $Q$ is the volume flow rate, $L$ is the packing's length in the flow direction, $A$ is the area of the packing's cross section, and $\Delta P$ is the pressure difference applied between the inlet and outlet surfaces of the aggregates packing.

## 3 Validation

### 3.1 X-CT

X-CT results are the foundation for the following numerical simulation of fluids flow in the aggregate packings. Thus, experiments on measuring the packing porosities were conducted to verify the accuracy of X-CT, and the experimental setup are shown in Fig. 3(a), where the volumetric cylinder and weighing scale are used to measure the volume and mass of infilled water in the packings, respectively. The porosity $\phi$ of the packing can be calculated by the following formular:

$$\phi = \frac{V_w}{V_c} = \frac{M_w}{\rho_w V_c}, \tag{14}$$

where $V_w$ is the volume of infilled water in the packing container, $V_c$ is the volume of the container, that is $100 \times 100 \times 100 = 10^6$ mm³, $M_w$ is the mass of infilled water, and $\rho_w$ is the density of water, that is 1000 kg/m³. According to Eq. (14), two methods based on the volume of water (VW) and mass of water (MW) were used independently in calculating the porosities of the packings. And the porosities measured by X-CT were elaborated in Sec. 2.2.

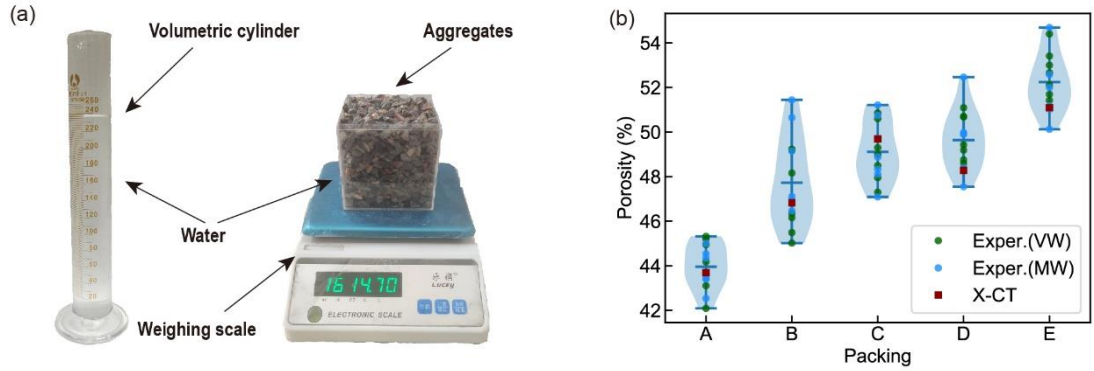

Figure 3 Validation of the X-CT method: (a) Experimental setup; (b) Comparison between the experiments and X-CT.

Each group of the aggregates had been packed randomly several times to measure the packing porosity by the above experimental setup, and the X-CT method was used for one packing for each group of the aggregates. Comparison results between the experiments based on volume of water (VM) and mass of water (MW) and X-CT are shown in Fig. 3(b) where violin plots were used to show the scattered data (i.e., porosities) for each packing. The width of the violin plot reflects the data distribution densities and the central bar represents the mean value of the data. Generally, the data measured by X-CT fall in the range of experimental results. This indicates the accuracy of the X-CT method is acceptable.

### 3.2 LBM

The accuracy of the LBM is also verified by experimental data [35] which were collected from experiments on permeability and porosity of Fontainebleau sandstone that is famous for its 100% quartz composition and large variation of porosities. The permeabilities were measured by a falling head method and the fluid used in the experiment was air. As the permeabilities used in this validation are larger than 10 mD, the Klingenberg effect was not considered here. For the porosity, it was measured by immersing the sample into toluene, and then the porosity can be calculated by the mass and density of toluene imbibed into the sample. And this method is also similar to that used in the validation of X-CT in Sec. 3.1.

3D digital models of Fontainebleau sandstone were also widely used in numerical simulations. The models in Ref. [36] had been compared and validated by experiments. Here in this work, four samples are used to validate the LBM in calculating the permeability as well as the porosity of the porous media. The digital data of each sample used here are $480 \times 480 \times 480$ voxels with a resolution of 5.7 $\mu$m/voxel. The four samples as shown in Fig. 4(a), 4(b), 4(c), 4(d) have a porosity around 8.4%, 9.8%, 12.2%, 17.6%, respectively. Water flow in the samples was simulated in the three directions of space, and thus three pairs of data points (porosity vs. permeability) of each sample were obtained. The initial and boundary conditions of the simulation were the same as those depicted in Sec. 2.4, that is the inlet and outlet surfaces of the sample were simulated by static pressure zones with a pressure difference of 0.1 Pa whereas the other surfaces were sealed. And the dynamic viscosity of water 0.001 Pa·s was used. After the simulation, the permeability and porosity of the sample can be calculated.

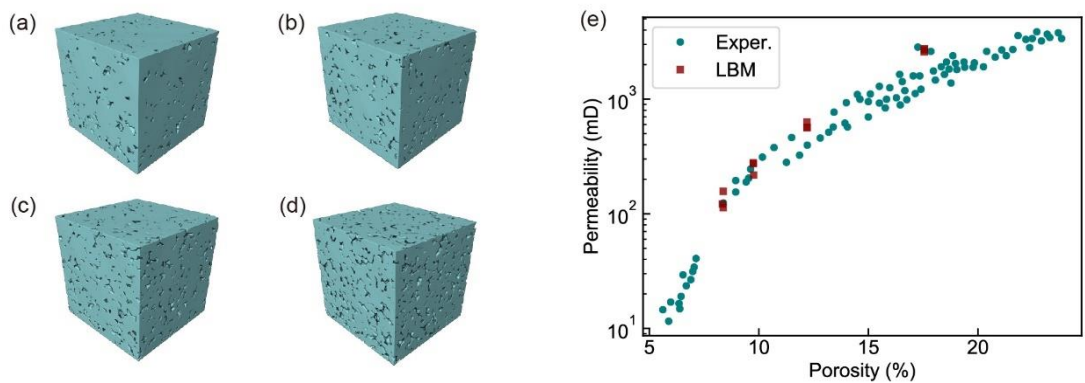

Figure 4 Validation of LBM in calculating the porosity and permeability of porous media: Digital Fontainebleau sandstone [36] with a porosity of (a) .8.4%; (b) 9.8%; (c) 12.2%; (d) 17.6%; (e) Comparison between experiments [35] and LBM.

Comparison between the experiments and LBM simulation is shown in Fig. 4(e), where 83 data points of experiments are shown in green circles and 12 data points of LBM are in purple squares. Some of the data points are overlapped and may not be differentiated. Generally, LBM data points agree well with experiments. Specifically, the LBM results of digital sample with higher porosity (e.g. 17.6%) may show a slightly higher permeability, but which is also in the range of experimental data. Especially, the digital Fontainebleau sandstone is actually known for this trend in comparison with the X-CT data of real sandstone [36]. Thus, the accuracy of the LBM can be higher when simulating the fluid flow in the X-CT reconstructed porous media.

## 4 Results

### 4.1 Porosities

The aggregates and pores of the packings together with their porosities obtained by the X-CT method are shown in Fig. 5, where all packings were cropped into cubes with $900 \times 900 \times 900$ voxels and a resolution of 0.098 mm/voxel. It can be found the aggregates in Figs. 5(a), 5(d), 5(g), 5(j), 5(m) show distinct size differences, and the pores of each packing, cf., Figs. 5(b), 5(e), 5(h), 5(k), 5(n), are connected with each other into a complicated network which occupies a large part of space in each packing.

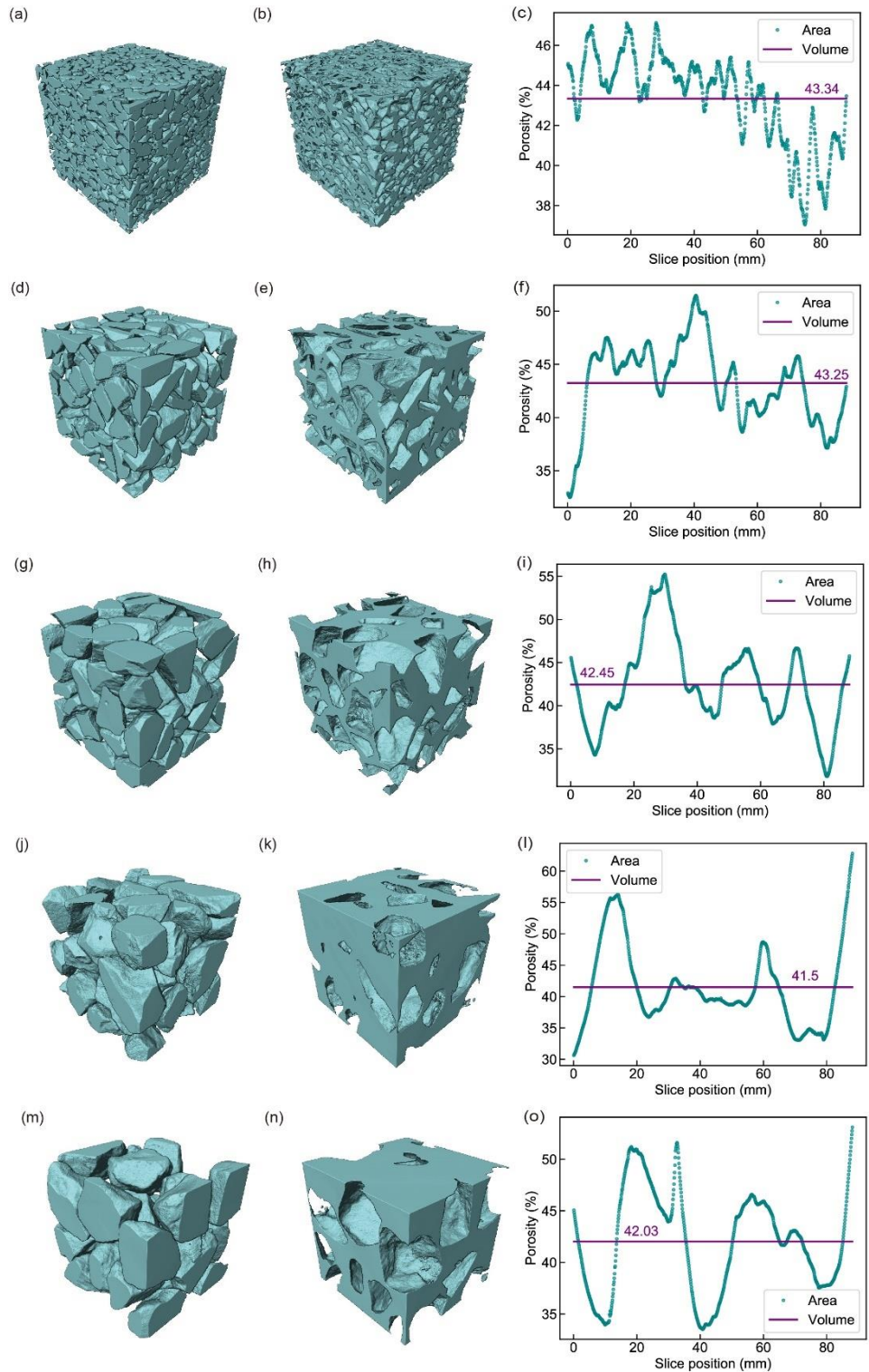

Figure 5. Aggregates, pores and porosity of packings (a, b, c) A, (d, e, f) B, (g, h, i) C, (j, k, l) D,

(m, n, o) E.

The porosities of each packing shown in Fig. 5 consist of two kinds of values, i.e., area porosity and volume porosity. The former is the pores' area fraction on each slice of the CT image, which are represented by green circles in Fig.5, and the latter represented by a straight purple line in Fig.5 is the pores' volume fraction of each packing. Generally, the area porosities show lots of fluctuations from slice to slice, whereas the volume porosities of each packing change little. Specifically, the maximum and minimum area porosities of packing A are 47.13% and 37.04%, respectively, which indeed show minimum change rate (i.e., 27.24%) among all packings, whereas the maximum change rate of area porosity (105.19%) occurred in packing D whose maximum and minimum area porosities are 62.81% and 30.61%, respectively. On the contrary, packing A has the maximum volume porosity which is 43.34% whereas packing D has the minimum volume porosity which is 41.5% and the change rate is merely 4.43%.

Nevertheless, the porosities in Fig. 5 are for the cropped specimens with a side length of 88.2 mm whereas the whole specimens were actually placed in a cubic container with a side length of 100 mm. The porosities of the whole packings were shown in Fig. 3(b) where it can be found the porosities of packings A to E with the container measured by X-CT are 43.696%, 46.827%, 49.69%, 48.281%, 51.092%, respectively. The differences between the porosities of cropped and uncropped specimens are largely due to the containers whose boundaries are different with the virtual cropped lines in the packings. This will be discussed in detail in Sec. 5.

**4.2 Pore networks**

Properties of pore networks are useful for the analyses of inner mass flow in the packings. Here the coordination numbers for the pores in the packings were calculated, which characterizes the connectivity of the pore networks. And the distributions of the coordination numbers are shown in Figs. 6(a)~6(e) for packings A to E, respectively. The data were fitted by Poisson distribution law. Generally, the fitting results are good except that of packing B whose coefficient of determination is relatively lower. This is because the randomness of the packing may lead to fluctuations. It should be noted that the maximum coordination number of the pore networks increases with the aggregate sizes. This is intuitively uncommon since the pores in the pore networks of larger aggregates, e.g., packing E, are relatively fewer, though they may be all well connected. The reason for this is the tiny pores of smaller scale in the larger-aggregate pore networks also exist but not intuitively noted from the illustrations in Fig. 5, and the greater pores of larger scale usually monopoly the majority connections of surrounding tiny pores. This feature may also facilitate the fluid flow in the pore networks. The fitting parameters $\lambda$ which is also the expectation of Poisson distribution law is shown in Fig. 6(f), where it can be found $\lambda$ of packing A has the largest value of 8.717 which indicates the pore in the network is expected to link with other 8.717 pores. From this point, pore network A can be deemed as well connected. But this is not to say the fluid in packing A flows faster than others, because the volume of pores may also influence the fluid flow.

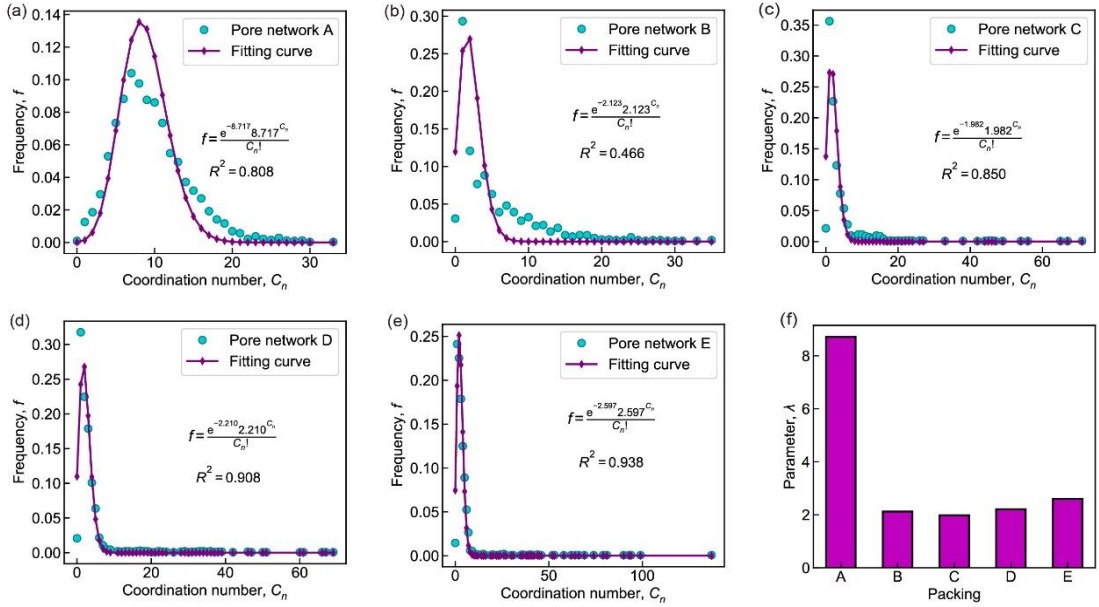

Figure 6. Distributions of coordination numbers for pore networks of packings (a) A, (b) B, (c) C, (d) D, (e) E, and their fitting parameters (f).

Possessing volume is a distinct feature of pore networks in comparison with social networks. Thus, the concept of coordination volume was proposed to reflect the volume-based connectivity of the pore networks. For the distribution of coordination volumes, results are shown in Figs. 7(a), 7(b), 7(c), 7(d), 7(e) for packings A, B, C, D, E, respectively. Power law was used to fit the distributions though all coefficients of determination are less than 0.9. This indicates the pore networks are not perfectly scale free, and thus the porosities may show clear fluctuations as illustrated in Fig. 5. However, the fitting parameters $\alpha$ and $\beta$ still have physical meanings and can reflect the physical properties of the pore networks. The parameter $\alpha$, which is the slope of the fitting line, shows a downward trend whereas the parameter $\beta$, which represents the intercept of the fitting line, shows upward trend with the increase of aggregate sizes as shown in Fig 7(f). From Figs 7(a) to 7(e), the fitting line rotates counterclockwise and thus lead to a decreasing $\alpha$. This also indicates the span of pores' volume increases. For the physical meaning of $\beta$, it represents the frequency of larger pores in the pore networks. As packing E has the largest $\beta$, that is to say larger pores in pore network E are relatively more. This can be confirmed by the illustration in Fig.5.

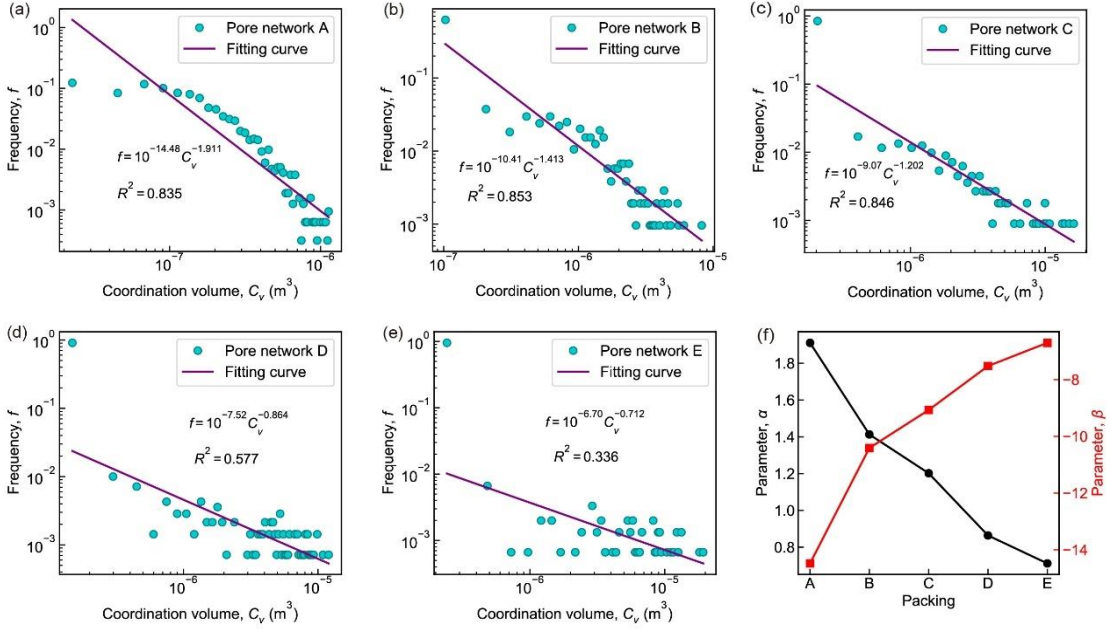

Figure 7. Distributions of coordination volumes for pore networks of packings (a) A, (b) B, (c) C, (d) D, (e) E, and their fitting parameters (f).

**4.3 Fluid flow**

Fluid (water) flow in the aggregates packing was simulated by the lattice Boltzmann method with BGK collision model in the pores' domain. The fluid velocity and mass flow rate were obtained and compared in Fig. 8 where the pore diameters calculated by the maximal-ball algorithm were also shown for reference. Generally, the pore diameter, fluid velocity and mass flow rate show upward trends with the increase of aggregate sizes.

From the sub-figures in the left column of Fig. 8, it can be found the maximum pore diameters of packings A, B, C, D, E are 7.37 mm, 15.18 mm, 21.77 mm, 22.99 mm, 35.60 mm, respectively. In comparison with the increment of aggregate sizes, the maximum pore diameter of packing D is not large enough, though it is also reasonable in view of the randomness of the packings. But this downward deviation leads to the maximum fluid velocity of packing D is slightly less than that of packing C, which, together with those of other packings, are shown in the central column of Fig. 8, where all the fluid flows are simulated from the top surface to the bottom surface of the cropped packings in three-dimensional space. The streamlines are shown in short white lines, which locate on the surface of fluids that are represented by the rainbow colors which also indicate the fluid velocity. From the velocity plots, it can be also found the larger velocity usually occurs in the narrow throats linking with broader pores. Such places are also the weak points for the mechanical strength of aggregates packing, and the flow impact force may increase the weakness of such places.

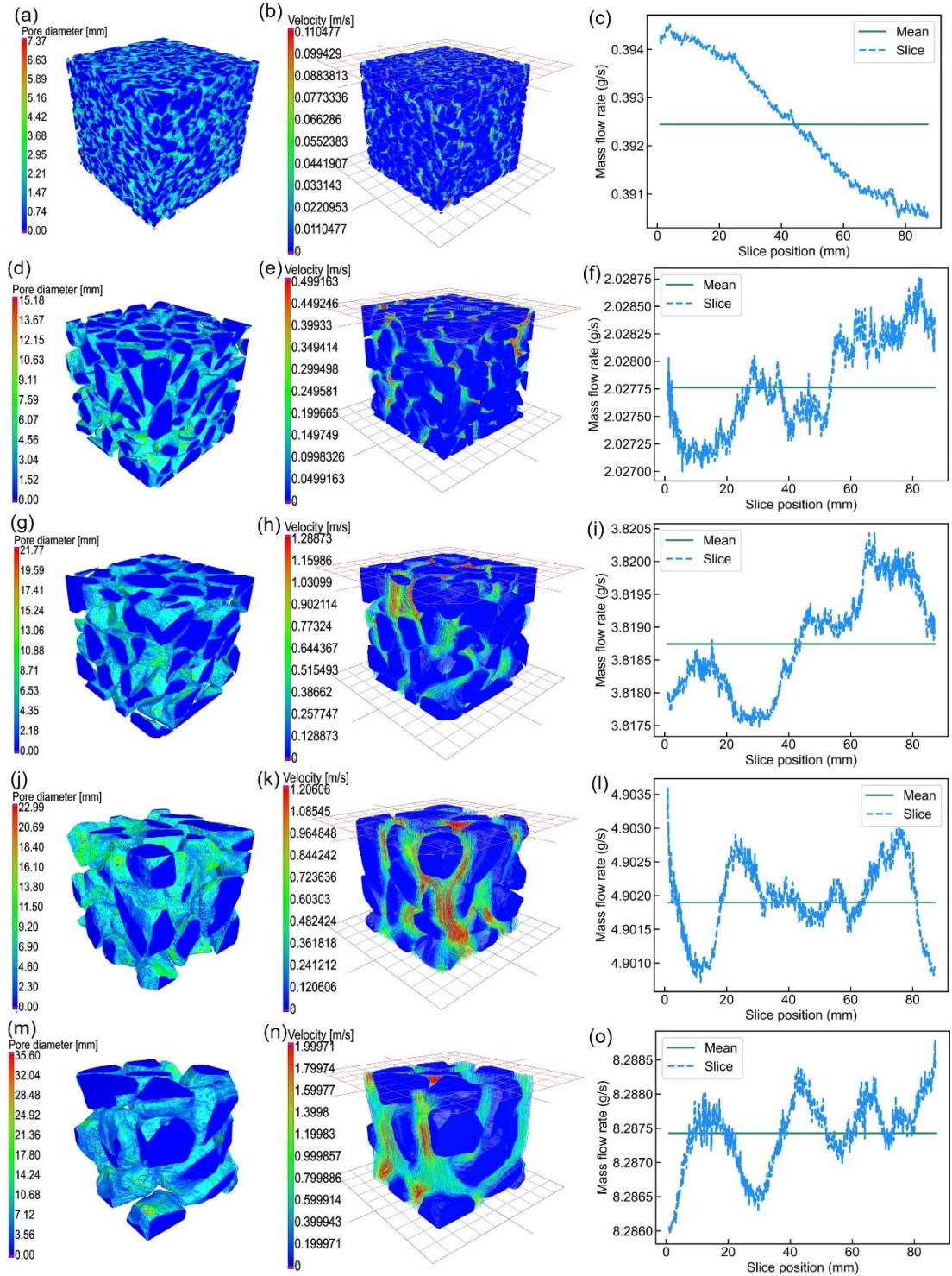

Figure 8. Pore diameter, fluid velocity and mass flow rate of packings (a, b, c) A, (d, e, f) B, (g, h, i) C, (j, k, l) D, (m, n, o) E, respectively.

The mass flow rate reflects the average fluid velocity through each slice. Results are shown in Figs. 8(c), 8(f), 8(i), 8(l), 8(o), respectively for the packings A, B, C, D, E, where the blue dash lines represent the mass flow rate of each slice (x-axis) and the green solid lines represent the average mass flow rates of the packings and their values from A to E are: 0.3924 g/s, 2.0278 g/s, 3.8187 g/s, 4.9019 g/s, 8.2874 g/s, respectively. Clearly, the average mass flow rates show an upward trend with the increase of aggregate sizes. For the mass flow rate on each slice, packing A shows a downward trend along the direction of fluid flow whereas the other packings show fluctuations due to the ratios of specimens' size to aggregate size are not as large as that of packing A. And in fact, the fluctuations also indicate the randomness of mass flow rate on each slice are remarkable.

Figure 9 shows the details of fluid velocity on specific slice of the packings. The slices, i.e., Figs. 9(a), 9(b), 9(c), 9(d), are chosen from packings A, B, D, E, respectively. The simulations are in three-dimensional space and indeed are the same as those in Fig.8. That is the fluids flow from the top to bottom surfaces of the packings whereas the other surfaces of the packings are sealed. In the same way, the streamlines are in white short lines which appear on the surface of fluids that are represented by rainbow colors which also indicates the fluid velocity. It can be found the sizes and connectivity of the void on the slice vary dramatically. This may provide knowledge for two dimensional simulations of fluids flow in aggregates packings in future. Secondly, the fluid velocity increases with the aggregate size. And the fluids generally find the shortest path to flow through the aggregates. In the path, the maximum velocity of the fluids on the referred slice usually appears in the narrow throat which links wider pores. This confirms the results found in three-dimensional velocities distributions in Fig. 8.

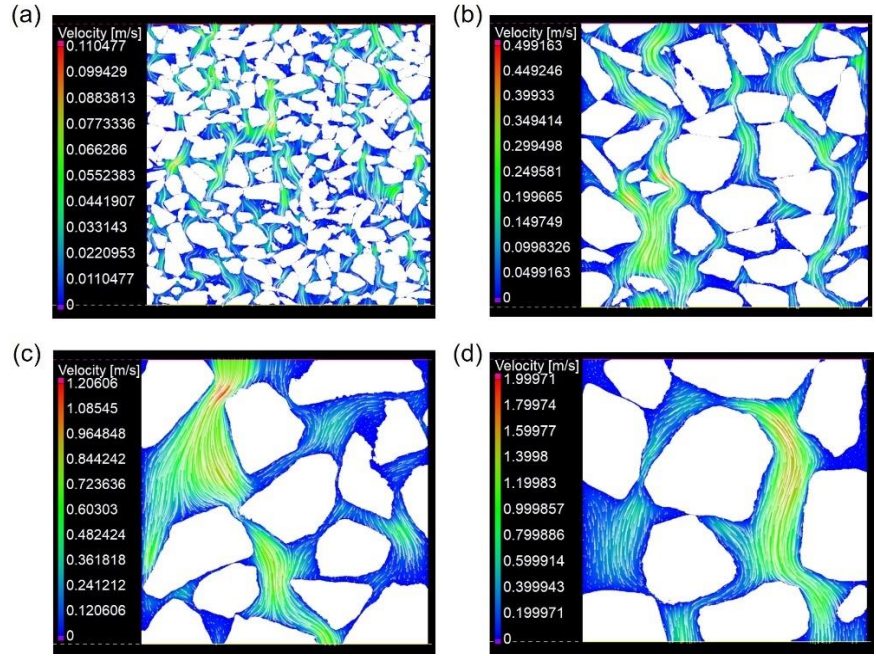

Figure 9. Fluid velocities on the slices of the aggregate packings (a) A, (b) B, (c) D, (d) E.

In addition, the average pore diameter, together with the permeability and tortuosity of the packings can be calculated during the fluid-flow simulation, and the results are shown in Fig. 10. It can be found the average pore diameter and permeability increase with the aggregate sizes whereas the tortuosity shows randomness when the aggregate size changes. Specifically, the average pore diameters for packings A, B, C, D, E are 2.45 mm, 5.81 mm, 8.67 mm, 10.16 mm, 13.81 mm, respectively. And the average permeability of the three directions of packings A, B, C, D, E are $0.553 \times 10^5$ D, $2.043 \times 10^5$ D, $6.056 \times 10^5$ D, $6.863 \times 10^5$ D, $9.429 \times 10^5$ D, $2.043 \times 10^5$ D, respectively. It shows clear upward trend, though the permeabilities of Y direction of packings B, C, D show slight fluctuations which are due to the randomness of the packings. Nevertheless, the average tortuosity of the three directions of packing A, B, C, D, E are 1.304, 1.302, 1.278, 1.302, 1.297, respectively, which are quite uniform in view of the changes of pore diameters and permeabilities. This indicates the tortuosity is less affected by the aggregate and pore sizes.

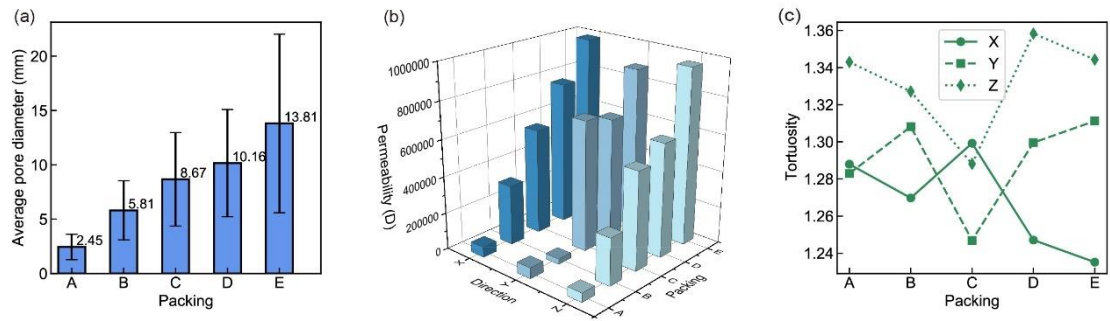

Figure 10. (a) Average pore diameter, (b) permeability and (c) tortuosity of the aggregate

packings.

The purpose of using the same size packings is to compare the fluid flow under the same condition. However, for the larger-size aggregates in packings B, C, D, E, the container with 10-cm side length limits the aggregates quantities. In fact, larger size packings had also been scanned by X-CT, but the X rays were attenuated significantly when they passed through the packings, though they were powered by a high-energy generator of 450 kV. The images reconstructed by the attenuated X-rays brought difficulties in segmentation and actually made it impossible by present technology. However, to form larger-size packings, another method exists, that is the reconstructed X-CT structures can be numerically merged several times. Here in this work, packings B, C, D, E were used to merge into larger-size packings. After merging, similar LBM simulations were conducted in the pore's domain of the packings, and the permeability, porosity of the packings as well as the fluid velocity were obtained and compared. Results are shown in Figs. 11 and 12.

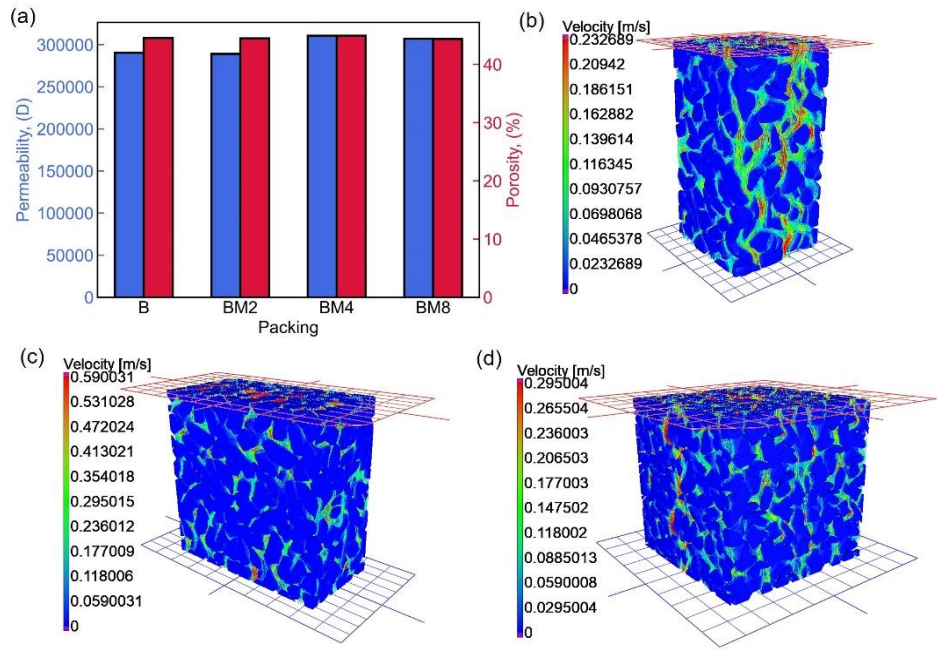

Figure 11. (a) Permeability and porosity of packings B, BM2, BM4, BM8; Fluid velocity of packings (b) BM2, (c) BM4, (d) BM8.

In Fig. 11, packings BM2, BM4, BM8 were formed by merging packing B twice, four times, eight times, respectively. The merging interfaces may bring unreality to the LBM simulation. Thus, increasing the merging times (to form larger-size packings) will enlarge the simulation error which can be reflected in the fluctuation of porosity. In Fig. 11(a), the porosities of packings B, BM2, BM4,

BM8 are 44.55%, 44.47%, 44.94%, 44.37%, respectively. The changes of porosities are trivial. Similar change trend occurs in permeability. But the fluid velocities, especially the maximum velocities, show distinct difference in Figs. 11(b), 11(c), 11(d) which are flow fields of packings BM2, BM4, BM8, respectively. Compared with the porosity changes in Fig 11(a), it can be found the maximum velocity of fluid increases with the increase of porosities. Therefore, packing BM4 has the largest maximum velocity of fluid among the packings in Fig. 11. On the other hand, the larger porosities of the merged packings indicate the merging interfaces are relatively more porous.

Similarly, packings CM3, DM4, EM5 were formed by merging packings C, D, E three times, four times, five times, respectively. Then, LBM simulations were conducted in the pores' domain of the merged packings. Results are shown in Fig.12. From Fig. 12(a), it can be found the porosity and permeability of the merged packings are close to those of the original packings, though packings DM4 and EM5 show relatively higher differences in comparison with their original packings D and E. This is because the merging times of such packings are relatively more. Similar to those of the original packings, the permeabilities of the merged packings also show upward trend with the increase of aggregate sizes, even though the porosities of the packings are close to each other. Moreover, among the six packings compared in Fig. 12(a), packing CM3 has the maximum porosity which is 45.18%. Accordingly, the highest maximum velocity exists in packing CM3 as shown in Figs. 12(b). Nevertheless, the maximum velocity not only depends on the porosity but also relates to the size of the packings. That is why the velocity fields of the merged packings show complicated fluctuations.

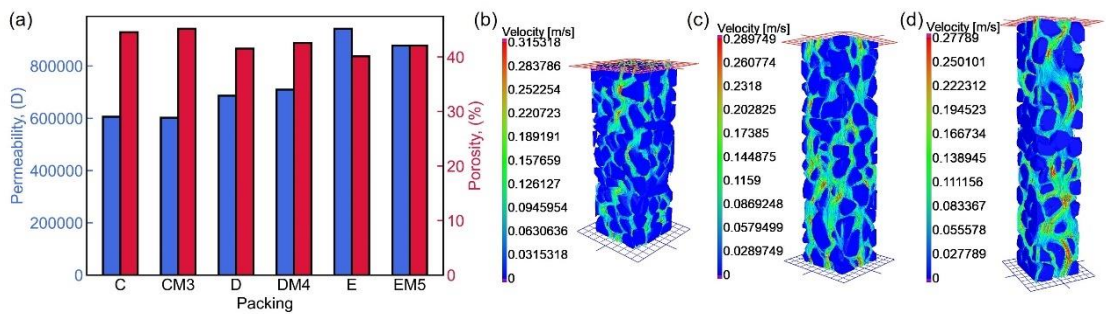

Figure 12. (a) Permeability and porosity of packings C, CM3, D, DM4, E, EM5; Fluid velocity of packings (b) CM3, (c) DM4, (d) EM5.

## 5 Discussion

The porosities of the cropped and uncropped packings show differences. For the uncropped whole packings in the cubic container, their average porosities show an upward trend with the increase of aggregate and pore sizes whereas those of cropped packings approximate to each other around 42.5%, with a maximum relative deviation of 4.43%. The reason for the differences is the wall effect as illustrated in Fig. 13, where the porosity measured around the container wall $\phi_b$ is usually greater than $\phi_i$ which is the porosity measured in the interior of the packing. The reason for this is the margin between the container and the packed aggregates increases the porosity and this increment becomes larger when the aggregates and pores are larger as the mechanism illustration shown in Fig. 13(b) where $\varepsilon_{b1}$, $\varepsilon_{b2}$ represent the errors brought by the wall effect for the porosities of larger-size-aggregate packing and smaller-size-aggregate packing, respectively. For the cropped packings, the wall effect is eliminated. This is also a merit of the X-CT method.

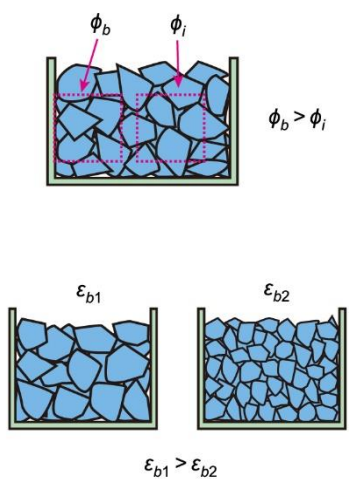

Figure 13. Wall effect of the container on the porosity test.

The fitting parameters of coordination number and coordination volume reflect the physical properties of the pore networks. As the two simple pore networks shown in Fig. 14, $\lambda$ represents the expectation of coordination number in the network. The connections among the pores of network #1 are more than those in network #2, thus, $\lambda_1 > \lambda_2$. Parameter $\alpha$ is the slope of the power-law fitting line in the logarithm coordinates system, which reflects the span of pores' volumes. Reviewing results in Fig. 7, it can be found $\alpha$ decreases when the pores' size span increases in a network. More straightforwardly, this is also illustrated in Fig.14 where $\alpha_1 > \alpha_2$ since the span of pores' volume in network #2 is relatively larger. Similarly, parameter $\beta$ represents the intercept of the fitting line and reflects the probability of finding a larger pore in a network. Thus, for the pore networks in Fig. 14, $\beta_1 < \beta_2$, as the frequency of larger pores in the pore network #2 is higher.

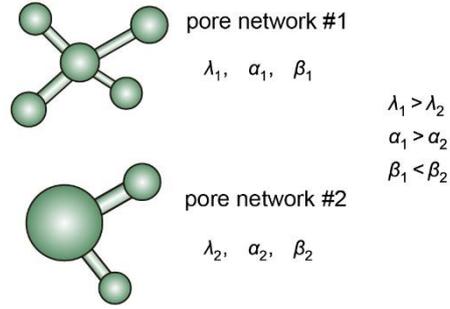

Figure 14. Comparisons of the fitting parameters of pore networks.

The mass flow rate together with the slice porosity is compared in Fig. 15 where the sub-figures (a), (b), (c), (d), (e) are for the packings A, B, C, D, E, respectively, and (f) and (g) are schematic illustrations for the relationships of mass flow rate and porosity of the two-dimensional slices and those of the three-dimensional blocks, respectively. Generally, for the two-dimensional slices, when the porosity $\phi$ goes up, the mass flow rate $\dot{m}$ decreases. For instance, in Fig. 15(f), when $\phi_1 > \phi_2$, $\dot{m}_1 < \dot{m}_2$. This is inferred from velocity distributions on each slice of the packings as shown in Fig. 9 where the fluids accelerate in narrow throats of the aggregates packing. However, for the three-dimensional packing block, when the volume porosity decreases, the mass flow rate shows a downward trend on the whole, as the illustration in Fig. 15(a) where $\phi_1 > \phi_2$, $\dot{m}_1 > \dot{m}_2$, that is also an opposite trend to the slice mass flow rate. The reason for this phenomenon is the fluid flows relatively faster when the void space is larger in the packing blocks since the flow resistance decreases.

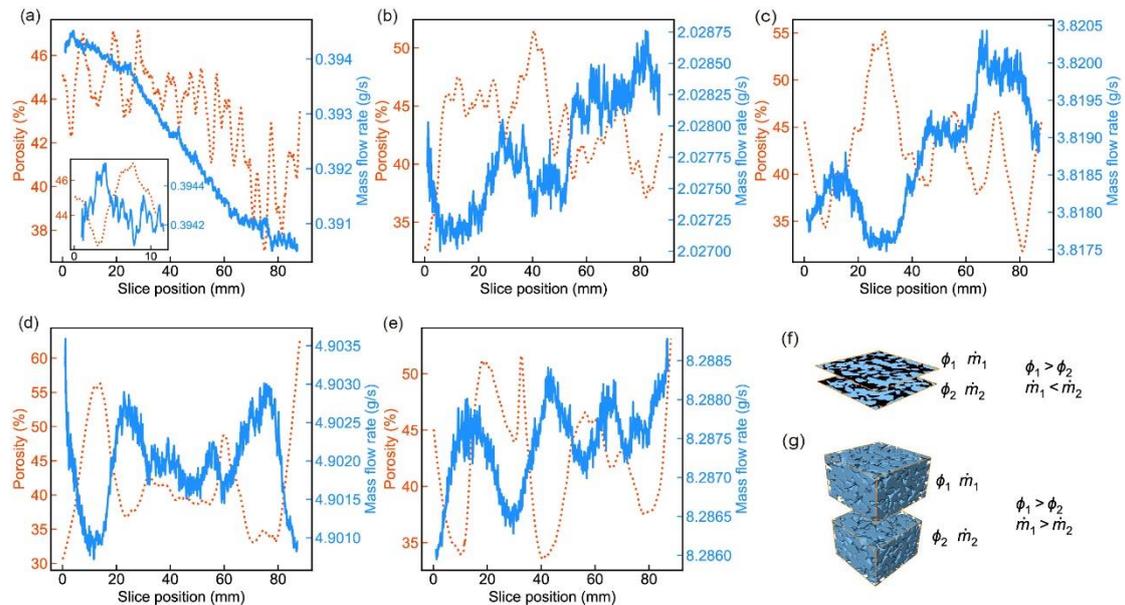

Figure 15. Comparisons of the mass flow rate and porosity of packings (a) A, (b) B, (c) C, (d) D,

(e) E, (f) two-dimensional slices, (g)three-dimensional blocks.

Permeability is an intrinsic property of the packing, which also reflects the average flow rate in the packing during fluids flow. Comparisons between the permeability with average pore diameter, porosity and tortuosity are shown in Figs. 16(a), 16(b) and 16(c), respectively. It can be found the permeability increases with the average pore diameter. This trend is consistent with previous studies [37, 38] where it is concluded that the packing permeability of idealized spherical particles (aggregates) is proportional to square of the particle's diameter. Nevertheless, particle shape also affects the permeability [39, 40]. In addition, according to Kozeny–Carman equation [8], permeability is also affected by the porosity and tortuosity. But the porosity and tortuosity are specifically uniform in this study. That is to say they may not show clear correlations with permeability as illustrated in Figs. 16(b) and 16(c). Hence, the changes of permeability are all reflected in the sizes and shapes of aggregates and pores of the packing, which are represented by the average pore diameters as illustrated in Fig. 16(a). This agrees with Kozeny–Carman equation [8].

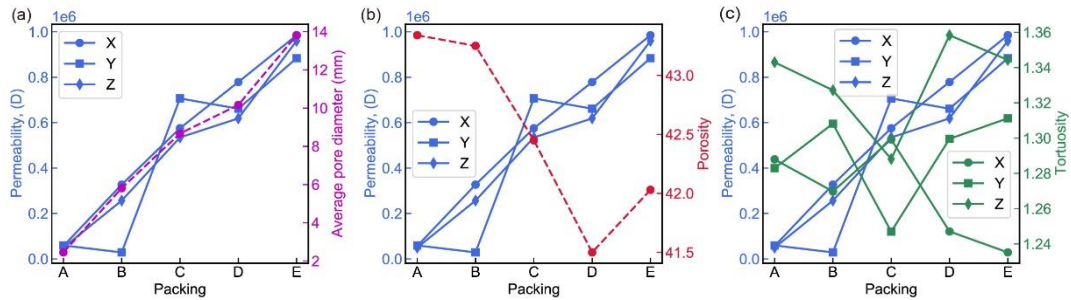

Figure 16. Comparisons between the permeability with average pore diameter, porosity and tortuosity of the aggregate packings.

Since the fitting parameters of coordination numbers and coordination volumes reflect the physical properties of the pore networks in the packing, the correlation between them and permeability is also of importance for understanding the fluid flow in specific pore networks. Hence, Fig. 17 shows the comparison results. Generally, the relationship between permeability and parameter $\lambda$ is unobvious whereas that between permeability with parameters $\alpha$ and $\beta$ is clearer. In Figs. 17(b) and 17(c), the permeability shows similar change trend with $\beta$ and opposite change trend with $\alpha$. From this point, it can be also found the volume properties of pore networks are more important and meaningful in comparison with the traditional properties related to social networks.

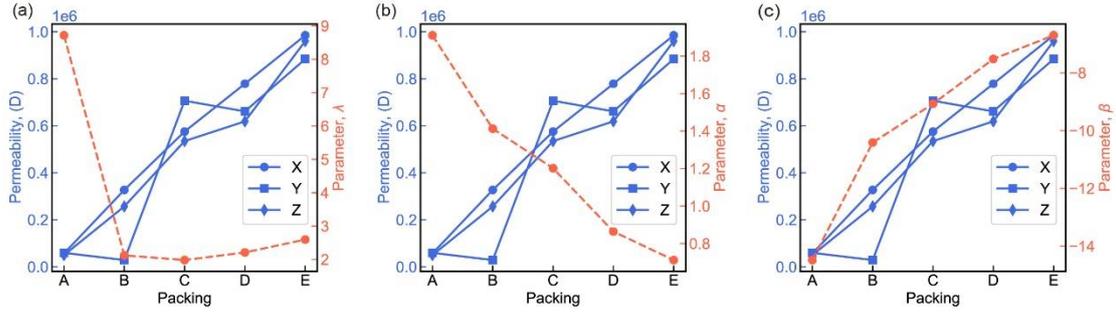

Figure 17. Comparisons between the permeability and the fitting parameters of pore networks.

The merged packings were used to show the stable macro parameters calculated by the present method, though the fluid velocities showed fluctuations, especially the maximum ones as shown in Figure 11. This is due to the discontinuity on the merging interfaces of the packings. And the error increases with the merging times. That is why merely limited merging times were used to form the larger merged packings. To illustrated this limitation, the average velocities of the fluid on each slice of the merged packings BM2, CM3, DM4, EM5 along their flow directions are shown in Figure 18 which also depicts the positions of the merging interfaces. It can be found the largest fluctuation of the average velocity usually occurred on or around the merging interfaces. And generally, the average velocity decreases significantly on the interfaces. This also indicates the fluid velocity in some pores around the merging interfaces may increase significantly, according to the fluid flow characteristics found in Figures 8 and 9. This brought unreality to the fluid flow on the interfaces, though it would not affect the macro parameters like bulk porosity and permeability calculated by fluid-flow simulations in the packings.

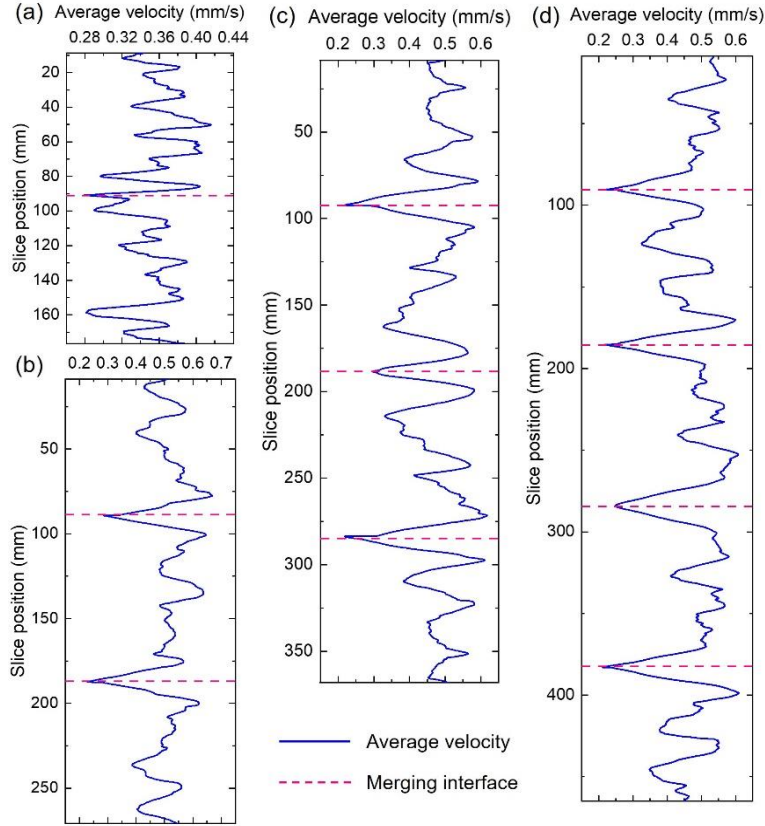

Figure 18. Average fluid velocities on the slices of merged packings (a) BM2, (b) CM3, (c) DM4, (d) EM5.

## 6 Conclusions

Fluids flow in the granular aggregates was investigated in this work. Conclusions are listed below:

- Wall effect on the porosity of aggregates packing is significant and the influence increases with the aggregate sizes. The pore sizes of the packing are positively proportional to the aggregate sizes whereas the porosity and tortuosity of the packing are relatively less influenced by the aggregate sizes.

- Poisson law and power law can be used to fit the coordination number and coordination volume of the packing's pore network, respectively. The fitting parameter of Poisson law ($\lambda$) reflects the connectivity of the pore networks, and the fitting parameters of power law ($\alpha$ and $\beta$) reflect the frequency of larger pores in the pore networks.

- Mass flow rates of fluids in the aggregates packing are affected by the porosities. On the two-dimensional slices, the mass flow rate decreases when the slice porosity increases. But for the three-dimensional packing blocks, the average mass flow rate increases with the volume porosity. And the permeability of the aggregates packing shows correlating change trend with the average pore diameter and fitting parameters of coordination volumes when the sizes of aggregates change.

# Acknowledgement

This work was supported by the National Natural Science Foundation of China (51908075).

# References

[1] Nemati KM, Uhlmeyer JS. Accelerated construction of urban intersections with Portland Cement Concrete Pavement (PCCP). Case Studies in Construction Materials. 2021;14:e00499.
[2] Zhang J, Ma G, Dai Z, Ming R, Cui X, She R. Numerical study on pore clogging mechanism in pervious pavements. Journal of Hydrology. 2018;565:589-98.
[3] Kusumawardani DM, Wong YD. Evaluation of aggregate gradation on aggregate packing in porous asphalt mixture (PAM) by 3D numerical modelling and laboratory measurements. Construction and Building Materials. 2020;246:118414.
[4] Takbiri-Borujeni A, Kazemi H, Nasrabadi N. A data-driven surrogate to image-based flow simulations in porous media. Computers & Fluids. 2020;201:104475.
[5] Kashefi A, Mukerji T. Point-cloud deep learning of porous media for permeability prediction. Physics of Fluids. 2021;33:097109.
[6] Reboucas RB, Loewenberg M. Resistance and mobility functions for the near-contact motion of permeable particles. Journal of Fluid Mechanics. 2022;938:A27.
[7] Sun Y, Causse P, Benmokrane B, Trochu F. Permeability Measurement of Granular Porous Materials by a Modified Falling-Head Method. Journal of Engineering Mechanics. 2020;146:04020101.
[8] Slotte PA, Berg CF, Khanamiri HH. Predicting Resistivity and Permeability of Porous Media Using Minkowski Functionals. Transport in Porous Media. 2020;131:705-22.
[9] Huang J, Xiao F, Dong H, Yin X. Diffusion tortuosity in complex porous media from pore-scale numerical simulations. Computers & Fluids. 2019;183:66-74.
[10] Gharibi F, Jafari S, Rahnama M, Khalili B, Jahanshahi Javaran E. Simulation of flow in granular porous media using combined Lattice Boltzmann Method and Smoothed Profile Method. Computers & Fluids. 2018;177:1-11.
[11] Gharedaghloo B, Price JS, Rezanezhad F, Quinton WL. Evaluating the hydraulic and transport properties of peat soil using pore network modeling and X-ray micro computed tomography. Journal of Hydrology. 2018;561:494-508.
[12] Martin T, Kamath A, Bihs H. A Lagrangian approach for the coupled simulation of fixed net structures in a Eulerian fluid model. Journal of Fluids and Structures. 2020;94:102962.
[13] Posa A, Broglia R, Balaras E. The dynamics of the tip and hub vortices shed by a propeller: Eulerian and Lagrangian approaches. Computers & Fluids. 2022;236:105313.
[14] Bahrami-Torabi H, Kerboua Y, Lakis AA. Finite element model to investigate the dynamic instability of rectangular plates subjected to supersonic airflow. Journal of Fluids and Structures. 2021;103:103267.
[15] Doi T. Effect of a small curvature of the surfaces on microscale lubrication of a gas for large Knudsen numbers. Physical Review Fluids. 2022;7:034201.
[16] Tian ZL, Liu YL, Zhang AM, Wang SP. Analysis of breaking and re-closure of a bubble near a free

surface based on the Eulerian finite element method. Computers & Fluids. 2018;170:41-52.

[17] Ye T, Pan D, Huang C, Liu M. Smoothed particle hydrodynamics (SPH) for complex fluid flows: Recent developments in methodology and applications. Physics of Fluids. 2019;31:011301.

[18] Mimeau C, Marié S, Mortazavi I. A comparison of semi-Lagrangian vortex method and lattice Boltzmann method for incompressible flows. Computers & Fluids. 2021;224:104946.

[19] Liu S, Jiang L, Wang C, Sun C. Lagrangian dynamics and heat transfer in porous-media convection. Journal of Fluid Mechanics. 2021;917:A32.

[20] Zhang S, Tang J, Wu H. Phase-field lattice Boltzmann model for two-phase flows with large density ratio. Physical Review E. 2022;105:015304.

[21] Ju L, Shan B, Yang Z, Guo Z. An exact non-equilibrium extrapolation scheme for pressure and velocity boundary conditions with large gradients in the lattice Boltzmann method. Computers & Fluids. 2021;231:105163.

[22] Visscher WM, Bolsterli M. Random Packing of Equal and Unequal Spheres in Two and Three Dimensions. Nature. 1972;239:504-7.

[23] Fu R, Hu X, Zhou B. Discrete element modeling of crushable sands considering realistic particle shape effect. Computers and Geotechnics. 2017;91:179-91.

[24] Wei D, Wang Z, Pereira J-M, Gan Y. Permeability of Uniformly Graded 3D Printed Granular Media. Geophysical Research Letters. 2021;48.

[25] Aljasmi A, Sahimi M. Speeding-up image-based simulation of two-phase flow in porous media with lattice-Boltzmann method using three-dimensional curvelet transforms. Physics of Fluids. 2021;33:113313.

[26] Lyu Q, Wu H, Li X. A 3D model reflecting the dynamic generating process of pore networks for geological porous media. Computers and Geotechnics. 2021;140:104444.

[27] Otsu N. A Threshold Selection Method from Gray-Level Histograms. IEEE Transactions on Systems, Man, and Cybernetics. 1979;9:62-6.

[28] Bultreys T, Lin Q, Gao Y, Raeini AQ, AlRatrout A, Bijeljic B, et al. Validation of model predictions of pore-scale fluid distributions during two-phase flow. Physical Review E. 2018;97:053104.

[29] Gackiewicz B, Lamorski K, Sławiński C, Hsu S-Y, Chang L-C. An intercomparison of the pore network to the Navier–Stokes modeling approach applied for saturated conductivity estimation from X-ray CT images. Scientific Reports. 2021;11:5859.

[30] Li G-y, Zhan L-t, Hu Z, Chen Y-m. Effects of particle gradation and geometry on the pore characteristics and water retention curves of granular soils: a combined DEM and PNM investigation. Granular Matter. 2021;23:9.

[31] Gostick J, Aghighi M, Hinebaugh J, Tranter T, Hoeh MA, Day H, et al. OpenPNM: A Pore Network Modeling Package. Computing in Science & Engineering. 2016;18:60-74.

[32] Lv Q, Liu X, Wang E, Wang S. Analytical solution to predicting gaseous mass flow rates of microchannels in a wide range of Knudsen numbers. Physical Review E. 2013;88:013007.

[33] Hafen N, Dittler A, Krause MJ. Simulation of particulate matter structure detachment from surfaces of wall-flow filters applying lattice Boltzmann methods. Computers & Fluids. 2022;239:105381.

[34] Lv Q, Wang E, Liu X, Wang S. Determining the intrinsic permeability of tight porous media based on bivelocity hydrodynetics. Microfluidics and Nanofluidics. 2014;16:841-8.

[35] Bourbie T, Zinszner B. Hydraulic and acoustic properties as a function of porosity in Fontainebleau Sandstone. Journal of Geophysical Research: Solid Earth. 1985;90:11524-32.

[36] Berg CF, Held R. Fundamental Transport Property Relations in Porous Media Incorporating Detailed

Pore Structure Description. Transport in Porous Media. 2016;112:467-87.

[37] Howells ID. Drag due to the motion of a Newtonian fluid through a sparse random array of small fixed rigid objects. Journal of Fluid Mechanics. 1974;64:449-76.

[38] Torquato SA, Haslach HW, Jr. Random Heterogeneous Materials: Microstructure and Macroscopic Properties. Applied Mechanics Reviews. 2002;55:B62-B3.

[39] Safari M, Gholami R, Jami M, Ananthan MA, Rahimi A, Khur WS. Developing a porosity-permeability relationship for ellipsoidal grains: A correction shape factor for Kozeny-Carman's equation. Journal of Petroleum Science and Engineering. 2021;205:108896.

[40] Bourbatache MK, Hellou M, Lominé F. Two-scale analysis of the permeability of 3D periodic granular and fibrous media. Acta Mechanica. 2019;230:3703-21.